\documentclass[12pt,a4paper,english,superscriptaddress,aps,nofootinbib]{revtex4}
\usepackage[utf8]{inputenc}
\usepackage[T1]{fontenc}
\usepackage{amsmath,amssymb,graphicx}
\makeatletter
\usepackage{babel}
\usepackage[active]{srcltx}
\usepackage{graphicx,color}
\usepackage{subfigure}
\usepackage{changebar}
\usepackage{upgreek}
\usepackage{hyperref}
\hypersetup{colorlinks=true,urlcolor= blue,citecolor=blue,linkcolor= blue}
\usepackage{braket}
\usepackage{dsfont}
\usepackage{mathtools}
\usepackage{slashed}
\usepackage{empheq}
\usepackage{tikz}
\usepackage{tikz-feynman}
\usepackage{multirow}
\usetikzlibrary{decorations.pathmorphing}

\usepackage{graphicx}
\usepackage{amsfonts}
\usepackage{amsmath}
\newcommand{\be}{\begin{equation}}
	\newcommand{\ee}{\end{equation}}
\newcommand{\bea}{\begin{eqnarray}}
	\newcommand{\eea}{\end{eqnarray}}

\begin{document}

\title{Topological and self-dual vortices in a double sigma model with Maxwell coupling}

\author{Francisco C. E. Lima}
\email{cleiton.estevao@ufabc.edu.br (Corresponding author)}
\affiliation{Centro de Matématica, Computação e Cognição (CMCC), Universidade Federal do ABC (UFABC), Av. dos Estados 5001, CEP 09210-580, Santo Andr\'{e}, S\~{a}o Paulo, Brazil.}

\author{Fernando M. Belchior}
\email{belchior@fisica.ufc.br}
\affiliation{Departamento de Física, Universidade Federal da Paraíba, 58051-970, João Pessoa, Paraíba, Brazil.}

\author{Allan R. P. Moreira}
\email{allan.moreira@fisica.ufc.br}
\affiliation{Secretaria da Educação do Ceará (SEDUC), Coordenadoria Regional de Desenvolvimento da Educação (CREDE 9), Horizonte, Ceará, 62880-384, Brazil.}

\begin{abstract}
In this work, we construct a double O(3)-sigma model minimally coupled to a Maxwell field in (2+1)-dimensional spacetime and investigate the existence of self-dual magnetic vortex solutions. An analysis of the Bogomol'nyi–Prasad–Sommerfield (BPS) property reveals that both sigma fields belong to the same topological sector and that the potential assumes a periodic cosine-like form. Furthermore, the theory supports the emergence of magnetic vortices with quantized flux, described by two nonlinear O(3)-sigma sectors that effectively combine into a single topological sector in the BPS regime. In addition, we analytically verify the consistency of the BPS structure and its asymptotic behavior. Within this framework, numerical vortex solutions confirm that the field profiles are smooth and spatially localized, with both the magnetic field and the energy density remaining regular and localized.
\end{abstract}

\maketitle


\section{Introduction}

Topological solitons play a central role across a wide range of areas in theoretical physics, from high-energy physics to condensed matter systems \cite{Bishop,Kosevich,Lima1,Lima2,Lima3}. In particular, these solutions appear in the context of topological defects in planar gauge theories, which have attracted considerable attention owing to their rich mathematical structure and their relevance to phenomena such as superconductivity \cite{Abrikosov1,Abrikosov2,Su,Kawana}, superfluidity \cite{Ewerz,Li}, and effective descriptions of three-dimensional systems \cite{Andrade,Bazeia}. One of the key features of these configurations is flux quantization, as well as their stability, which is a consequence of topological boundary conditions \cite{Nielsen}.

Among the various scenarios that support vortex solutions, nonlinear sigma models provide a natural framework for exploring the interplay between geometry and topology \cite{Schroers,Leese,Ghosh,Ferko}. In particular, O(3)-sigma models have been extensively studied as effective descriptions of systems whose internal degrees of freedom are constrained to a spherical target, exhibiting nontrivial mappings associated with topological sectors \cite{Bruckmann}. Upon coupling to a gauge field, the NLSMs acquire an even richer structure, which allows the emergence of gauged solitons whose properties are determined by both the target space and the underlying gauge field \cite{Cunha,Lima4}.

Within this framework, a particularly relevant aspect is the study of self-dual configurations, known as Bogomol’nyi-Prasad-Sommerfield (BPS) solutions. These solutions arise when the system satisfies the BPS condition, which allows for a reduction in the order of the equations of motion while ensuring their stability, see Ref. \cite{Bogomolnyi,Prasad}. Owing to this property, the BPS structure has been successfully applied in a variety of contexts \cite{Canfora,Izquierdo,Alonso}, including the Abelian Higgs model \cite{DBazeia}, Chern–Simons theory \cite{Kim}, and gauged sigma models \cite{Lima6,Lima7}, revealing deep connections between topology, energy minimization, and integrability.

Recently, there has been growing interest in systems with multiple interacting topological sectors \cite{Matheus,Lima8}, motivated by the need to describe more complex physical scenarios. In this framework, sigma models with multiple fields coupled to a single gauge field can provide a fertile medium for investigating the emergence of internal structures and their influence on the formation and properties of such topological defects. Similarly, the presence of multiple sigma sectors may induce nontrivial cooperative or competitive effects, modifying the vortex core structure, the energy distribution, and the behavior of the magnetic flux.

In this work, we investigate a double O(3)-sigma model minimally coupled to the Maxwell field in a $(2+1)$-dimensional spacetime. Our main purpose is to examine the existence of self-dual magnetic vortex solutions in a system with two independent sigma sectors that interact through a single gauge field. We show that the consistency of the BPS property imposes strong constraints on the model, leading to a symmetric self-dual sector in which both sigma fields effectively collapse into a single topological configuration.

We organized this work into six sections. Section \ref{SecII} introduces the theoretical model and derives the corresponding equations of motion. Section \ref{SecIII} discusses the vortex ansatz and its topological properties. In Section \ref{SecIV}, we examine the BPS configurations. Section \ref{SecV} provides numerical solutions and analyzes their physical behavior. Finally, Section \ref{SecVI} summarizes the main findings.


\section{Theoretical framework}\label{SecII}

The simultaneous presence of two O(3)-sigma fields, $\Phi$ and $\Theta$, coupled to the same gauge field, allows one to investigate how distinct sectors may coexist, compete, and/or cooperate in the formation of vortex-like structures, yielding a richer dynamical behavior than that noted in models governed by a single scalar field. From a mathematical viewpoint, the model provides a natural framework for analyzing flux quantization, topological structure, and stability. From a physical perspective, it can serve as an effective laboratory for describing systems with multiple order parameters and coupled sectors, in which topological defects play a central role. In this context, the study of this model helps to clarify, in a controlled manner, how gauge interactions and the topological sectors inherited from the O(3)-sigma model influence the physical properties of topological vortices.

To accomplish our purpose, let us begin by considering the most general Lagrangian density for two $O(3)$-sigma fields coupled to the Maxwell field in a flat three-dimensional spacetime\footnote{Within this theoretical framework, we consider a $(2+1)$-dimensional spacetime with metric signature $g_{\mu\nu}=\mathrm{diag}(-,+,+)$. We further emphasize that the O(3)-sigma fields $\Phi$ and $\Theta$ are scalar triplets subject to the constraints $\Phi \cdot \Phi = 1$ and $\Theta \cdot \Theta = 1$.}, viz., 
\begin{align}\nonumber
    \mathcal{S}=&\int\,d^3x\,\Bigg[\frac{\mathcal{F}(\Theta)}{2}D_{\mu}\Phi\cdot D^{\mu}\Phi+\frac{1}{2}D_{\mu}\Theta\cdot D^{\mu}\Theta-\frac{1}{4}F_{\mu\nu}F^{\mu\nu}+\lambda_\Phi(\Phi\cdot\Phi-1)+\lambda_\Theta(\Theta\cdot\Theta-1)\\
    -&V(\Phi,\Theta)\Bigg],
\label{Eq1}
\end{align}
where $\Phi$ and $\Theta$ are O(3)-sigma fields, $V(\Phi,\Theta)$ is the potential, and $\lambda_{\Phi}$ and $\lambda_{\Theta}$ are Lagrange multipliers. Meanwhile, $D_\Phi$ and $D_\mu\Theta$ are, respectively, the covariant derivatives, namely,
\begin{align}
D_{\mu}\Phi=\partial_{\mu}\Phi-eA_{\mu}(\hat n_{3}\times\Phi)
\qquad \mathrm{and} \qquad
D_{\mu}\Theta=\partial_{\mu}\Theta-eA_{\mu}(\hat n_{3}\times\Theta),
\label{Eq2}
\end{align}
with $F_{\mu\nu} = \partial_\mu A_\nu - \partial_\nu A_\mu$, i.e., the electromagnetic field tensor, $e$ is the charge and $\hat{n}_3$ is the unit vector $\hat{n}_3 = (0,0,1)$.

Having established the theoretical model, let us examine the equations of motion corresponding to the action [Eq. \eqref{Eq1}]. To derive them, we apply the principle of least action by varying the action with respect to the fields $\Phi$, $\Theta$, and $A_\mu$, which yields 
\begin{align}
    &D_\mu\left(\mathcal{F} D^\mu\Phi\right)+V_\Phi=\lambda_\Phi\Phi, \qquad D_\mu D^\mu \Theta+\frac{1}{2}\mathcal{F}_\Theta (D_\mu\Phi\cdot D^\mu\Phi)+V_\Theta=\lambda_\Theta \Theta,
\label{Eq3}\end{align}
and
\begin{align}
    \partial_\mu F^{\mu\nu}=J^\nu \qquad \mathrm{with} \qquad J^\nu=-e\hat{n}_3[\mathcal{F}\,(\Phi\times D^\nu\Phi)+\Theta\times D^\nu\Theta],
    \label{Eq4}
\end{align}
where $\mathcal{F}_\Theta=\partial\mathcal{F}/\partial\Theta$, $V_\Phi=\partial V/\partial \Phi$, and $V_\Theta=\partial V/\partial\Theta$. Furthermore, $\lambda_\Phi$ and $\lambda_\Theta$ are the Lagrange multipliers, viz.,
\begin{align}
    \lambda_\Theta=-D_\mu\Phi\cdot D^\mu\Phi+\frac{1}{2}(\Theta\cdot \mathcal{F}_\Theta)(D_\mu\Phi\cdot D^\mu\Phi)+\Theta\cdot V_\Theta,
    \label{Eq5}
\end{align}
and
\begin{align}
    \lambda_\Phi=-\mathcal{F}\,D_\mu\Phi\cdot D^\mu\Phi+\Phi\cdot V_\Phi.
    \label{Eq6}
\end{align}
Therefore, the equations of motion are given by:
\begin{align}
    \mathcal{F}\,D_\mu D^\mu\Phi+(\partial_\mu\mathcal{F})D^\mu\Phi+\mathcal{F}(D_\mu\Phi\cdot D^\mu\Phi)\cdot\Phi+V_\Phi-\left(\Phi\cdot V_\Phi\right)\Phi=0,
    \label{Eq7}
\end{align}
\begin{align}
    D_\mu D^\mu\Theta+(D_\mu\Theta\cdot D^\mu\Theta)\Theta+\frac{1}{2}\left[\mathcal{F}_\Theta-(\Theta\cdot \mathcal{F}_\Theta)\Theta\right](D_\mu\Phi\cdot D^\mu\Phi)+V_\Theta-\left(\Theta\cdot V_\Theta\right)\Theta=0,
    \label{Eq8}
\end{align}
and 
\begin{align}
    \partial_\mu F^{\mu\nu}=J^\nu \qquad \mathrm{with} \qquad J^\nu=-e\hat{n}_3[\mathcal{F}\,(\Phi\times D^\nu\Phi)+\Theta\times D^\nu\Theta].
    \label{Eq9}
\end{align}
One highlights that the current $J^\nu$ acts as a dynamical source for the gauge field, encoding the coupling between the two sigma sectors. In particular, $\Phi$ and $\Theta$ respond to the $U(1)$ symmetry associated with internal rotations around the $\hat{n}_3$ axis. Accordingly, the current measures the charged content of the transverse components of the fields, i.e., the parts of $\Phi$ and $\Theta$ that lie in the target space orthogonal to $\hat{n}_3$, and therefore it directly governs the resulting electromagnetic configuration. The Maxwell equation, $\partial_\mu F^{\mu\nu} = J^\nu$, makes it clear that the gauge field is neither external nor passive. Conversely, within the sigma sectors, the presence of $\mathcal{F}(\Theta)$ simultaneously modifies the dynamics of the field $\Phi$, the back-reaction of the $\Phi$ sector on the field $\Theta$, and the source. Furthermore, one notes that the $\Theta$ sector locally controls the strength of the kinetic term of $\Phi$.


\section{On the physical properties of the vortex}\label{SecIII}

Since the sigma fields exhibit an angular dependence proportional to the winding number $n$, the ordinary derivative $\partial_\theta$ generates a contribution that would, by itself, lead to a divergence of the energy at spatial infinity \cite{Vachaspati,MantonS}. To solve this issue, it is necessary to implement an axially symmetric gauge field ($A_r=0$) to bypass this angular variation. Therefore, let us implement an axially symmetric ansatz \cite{Nielsen}, i.e.,
\begin{align}
    A_\theta=\frac{n-a(r)}{er},
    \label{Eq10}
\end{align}
where $n$ is the winding number. In this case, the parametrization in terms of the field variable $a(r)$ proves convenient, as it clearly separates the topological contribution, governed by $n$, from the dynamical content encoded in the radial profile $a(r)$. Naturally, for the ansatz \eqref{Eq10} the magnetic field is oriented along the vortex axis and localized near the core. Moreover, imposing the asymptotic condition $a(\infty)\to\beta_\infty$ ensures the proper compensation of the angular phase at infinity and naturally leads to the quantization of the magnetic flux.

For the topological sector of the O(3)-sigma fields \cite{Rajaraman}, one adopts
\begin{align}
    \Phi=\begin{pmatrix} \sin f(r)\cos n\theta\\ \sin f(r)\sin n\theta\\ \cos f(r)\end{pmatrix}\qquad \mathrm{and} \qquad \Theta=\begin{pmatrix}\sin g(r)\cos n\bar\theta\\ \sin g(r)\sin n\bar\theta\\ \cos g(r) \end{pmatrix}.
    \label{Eq11}
\end{align}

Naturally, for the ansatz announced in Eq. \eqref{Eq10}, one obtains the magnetic field $B=F_{12}=-\frac{a'(r)}{e\,r}$, which allows us to conclude that our vortex configuration will have a topological magnetic flux given by
\begin{align}
    \varphi_{\mathrm{flux}}=\int\, B\,d^2x=\frac{2\pi}{e}[a(0)-a(\infty)].
    \label{Eq12}
\end{align}
Therefore, one notes that the topological character of vortex solutions is manifest, once the magnetic flux is independent of the local details of the field profiles and depends solely on the boundary values of the field variable $a(r)$. This result demonstrates that the flux is a topologically protected quantity, invariant under continuous deformations of the solution, reflecting the fact that the gauge field globally compensates for the angular variation of the sigma fields. Physically, this implies that the vortex carries a magnetic flux confined to its core, defined purely by the topological boundary conditions. 

\subsection{The topological boundary conditions}

For the theory introduced in Eq. \eqref{Eq1}, let us assume the topological conditions, viz.,
\begin{align}
    f(0)=0, \quad g(0)=0, \qquad a(0)=n,\quad
    f(\infty)=\pi, \quad g(\infty)=\pi, \quad a(\infty)=\beta_\infty.
    \label{Eq13}
\end{align}
Here, $\beta_\infty \in \mathbb{Z}^+$ and $n$ is the winding number. The above conditions encode a configuration that is both topologically nontrivial and energetically admissible. At the core, $f(0)=g(0)=0$ implies that both sigma sectors point to the north pole in the internal space $(\Phi=\Theta=0)$, ensuring regularity and the absence of singularities at the origin. At spatial infinity, $f(\infty)=g(\infty)=\pi$ enforces that the fields approach a vacuum configuration, so that the solution interpolates between distinct points on the target manifold $S^2$, thereby characterizing a nontrivial topological sector. The boundary conditions for the gauge field, $a(0)=n$ and $a(\infty)=\beta_\infty$, with $\beta_\infty \in \mathbb{Z}$, reflect the requirement of finite energy. Particularly, at infinity the covariant derivative must cancel the angular variation of the fields, which fixes the asymptotic value of $A_\theta$, and consequently of $a(r)$, thereby ensuring a quantized magnetic flux, i.e., 
\begin{align}
    \varphi_{\mathrm{flux}}=\frac{2\pi M}{e},
    \label{Eq14}
\end{align}
where $M=n-\beta_\infty\in\mathbb{Z}$ and $\beta_\infty=1,2,3,\cdots$. It follows that $n>\beta_\infty$, and the magnetic flux is quantized by the integer $M \in \mathbb{Z}$.


\section{On the BPS property}\label{SecIV}

Let us examine the BPS property associated with the theory announced in Eq. \eqref{Eq1}. In this case, upon adopting the metric signature, the energy density takes the form $\mathcal{E}=-\mathcal{L}$. Therefore, by considering Eqs. \eqref{Eq10} and \eqref{Eq11}, the total energy will be \footnote{Within this theoretical framework, we adopt the natural units, i.e., $\hbar=e=c=1$. Furthermore, we recall that the metric signature is $g_{\mu\nu}=\mathrm{diag}(+,-,-)$.}
\begin{align}
    \mathrm{E}=\int\,d^2x\,\left[\frac{\mathcal{F}(g)}{2}f'^2+\frac{1}{2}g'^2+\frac{\mathcal{F}a^2}{2r^2}\sin^2(f)+\frac{a^2}{2r^2}\sin^2(g)+\frac{a'^2}{2r^2}+V\right],
    \label{Eq15}
\end{align}
i.e., 
\begin{align}
    \mathrm{E}=2\pi\int_{0}^{\infty}\, r dr\,\left[\frac{\mathcal{F}(g)}{2}\left(f'^2+\frac{a^2}{r^2}\sin^2(f)\right)+\frac{1}{2}\left(g'^2+\frac{a^2}{r^2}\sin^2(g)\right)+\frac{a'^2}{2r^2}+V(f,g)\right].
    \label{Eq16}
\end{align}

Upon introducing the superpotential $\mathcal{W}$ \footnote{One highlights that the superpotential is an auxiliary scalar function $\mathcal{W}(\Phi,\Theta)$ that allows us to reformulate the potential in a factorized form, typically as a sum of squares of derivatives of $\mathcal{W}$. This construction is not merely formal; rather, it is related to the possibility of reorganizing the static energy via the completing-the-square procedure, leading to a Bogomol’nyi bound that is saturated by BPS solutions. The introduction of the superpotential leads us to a substantial reduction in dynamical complexity, replacing second-order equations of motion with a first-order BPS system, and ensuring its topological stability. For further details on the superpotential function, see Refs. \cite{Vachaspati}.}, it follows that 
\begin{align}\nonumber
    \mathrm{E}=&2\pi\,\int_{0}^{\infty}\,r dr\,\Bigg[\frac{\mathcal{F}}{2}\left(f'\pm\frac{a}{r}\sin f\right)^2+\frac{1}{2}\left(g'\pm\frac{a}{r}\sin g\right)^2+\frac{1}{2}\left(\frac{a'}{r}\mp \mathcal{W}\right)^2+\left(V-\frac{1}{2}\mathcal{W}^2\right)+\\
    &\pm\frac{1}{r}\frac{d}{dr}(a\mathcal{W})\mp \frac{a}{r}\mathcal{W}'\mp\frac{a}{r}(\mathcal{F}f'\sin f+g'\sin g)\Bigg].
    \label{Eq17}
\end{align}
Therefore, for the model to have the BPS property, one must require $\mathcal{F}(g)=1$. Furthermore, one requires that
\begin{align}
    V=\frac{1}{2}\mathcal{W}^2
    \label{Eq18}
\end{align} 
with
\begin{align}
    \mathcal{W}=\cos f+\cos g+c_0.
    \label{Eq18}
\end{align}
Here, $c_0$ is an integration constant, which we chosen as $c_0=-1$ \footnote{The choice $c_0=1$ is adopted to ensure that the spontaneous symmetry breaking is omnipresent.}. Naturally, this choice leads to $\mathcal{W}=\cos f+\cos g-1$.

Therefore, the energy \eqref{Eq18} boils down to
\small{\begin{align}
    \mathrm{E}=&2\pi\,\int_{0}^{\infty}\,r dr\,\Bigg[\frac{\mathcal{F}}{2}\left(f'\pm\frac{a}{r}\sin f\right)^2+\frac{1}{2}\left(g'\pm\frac{a}{r}\sin g\right)^2+\frac{1}{2}\left(\frac{a'}{r}\pm \mathcal{W}\right)^2\Bigg]+\mathrm{E}_{\mathrm{BPS}},    \label{Eq19}
\end{align}}
where
\begin{align}
    \mathrm{E}_{\mathrm{BPS}}=\pm 2\pi\int_{0}^{\infty}\, \frac{d}{dr}(a\mathcal{W})\, dr=\mp\, 4\pi(\beta_\infty+n).
    \label{Eq20}
\end{align}

Note that Eq. \eqref{Eq20} implies that $\mathrm{E}\geq\mathrm{E}_{\mathrm{BPS}}$, i.e., the energy is bounded from below. Therefore, in the saturation limit, one obtains the self-dual equations, viz., 
\begin{align}
    f'=\mp\frac{a}{r}\sin f, \qquad g'=\mp\frac{a}{r}\sin g, \qquad a'=\pm\,r\,(\cos f+\cos g-1),
    \label{Eq21}
\end{align}
which allows us to conclude that $f=g$. Accordingly, the system of BPS equations reduces to \begin{align}
    f'=\mp\frac{a}{r}\sin f \qquad \mathrm{and} \qquad a'=\pm r\,(2\cos f-1).
    \label{Eq22}
\end{align}
The form of the BPS equations [\eqref{Eq21}–\eqref{Eq22}] makes explicit the profile of the effective physical structure in the self-dual regime. In this framework, the dynamics of the sigma fields become governed by a single scalar degree of freedom $f(r)$, coupled to the gauge field $a(r)$, reflecting a merging of the two sigma sectors into a single topological sector. This result is not merely a mathematical simplification; rather, it indicates that, in the energy-saturating limit, the two originally independent fields become indistinguishable, jointly giving rise to the magnetic vortex structure.

\subsection{The asymptotic behaviors}

Allows us to analyze the asymptotic behavior of the topological solutions near the vortex core, i.e., when $r \to 0$. In this regime, the self-dual equation for the gauge field leads to
\begin{align}
    a'\simeq\mp 2r,
    \label{Eq23}
\end{align}
which leads to
\begin{align}
    a\simeq n\mp r^2+\cdots \qquad \mathrm{when} \qquad r\to 0.
    \label{Eq24}
\end{align}

Meanwhile, for the topological sector of the sigma field $f(r)$, one obtains
\begin{align}
    f\simeq f_0\,r^{\vert n\vert} \qquad \mathrm{when} \qquad r\to 0.
    \label{Eq25}
\end{align}
where $f_0$ is an integration constant. Note that, at the core, the fields exhibit regular behavior with $a(0)=n$ and $f(0)=0$. In this limit, one remarks that the winding number $n$ directly controls the order of vanishing of the sigma field at the vortex center. Accordingly, the azimuthal component of the gauge potential compensates the angular dependence of the sigma sector, ensuring both finite energy and regularity at the core. Meanwhile, the power-law behavior of $f(r)$ characterizes the formation of the topological vortex. Specifically, the dependence $f(r)\propto r^{\vert n\vert}$ indicates that higher-winding vortices exhibit a more flattened core, i.e., the approach to the internal vacuum becomes slower near the core. These results confirm that the physical structure of the defect is governed by the interplay between topology, gauge coupling, and the internal geometry, with the local behavior at the core remaining fully consistent with the existence of finite-energy BPS solutions.

We now examine the regime far from the vortex, i.e., as $r \to \infty$. In this case, one considers perturbations around the asymptotic values, namely, $f \approx \pi - \eta$ and $a(r) = \beta_\infty + \delta a(r)$, with $\eta \ll 1$ and $\delta a \ll 1$, and $\eta(0)=0$, yielding 
\begin{align}
    a(r)\sim \beta_\infty\mp\frac{3}{2}r^2+\cdots \qquad \mathrm{when} \qquad r\to \infty,
    \label{Eq26}
\end{align}
which implies a quadratic growth of the gauge field at infinity. Therefore, for the model to support regular solutions, one must require $\beta_\infty = 0$. Hence, the asymptotic condition $\beta_\infty = 0$ is not an arbitrary choice, but a necessary consequence of the internal consistency of the linearized BPS equations at spatial infinity.

To conclude the asymptotic analysis, we examine the behavior of the field $f(r)$, which yields
\begin{align}
    f(r)\simeq \pi - \bar{f}_\infty\,r^{-\beta_\infty}+\cdots \qquad \mathrm{when} \qquad r\to \infty,
    \label{Eq27}
\end{align}
where $\bar{f}_\infty$ is an arbitrary integration constant. The asymptotic expression obtained for $f(r)$ shows that the scalar field approaches its vacuum value $\bar{f}_\infty = \pi$ in an polynomial damped manner. This behavior indicates that the solution is spatially localized, with perturbations around the vacuum being rapidly suppressed. Furthermore, this polynomial decay implies that the contribution of the scalar field to the energy density becomes negligible at infinity, ensuring that the configuration quickly approaches the vacuum sector and that the BPS solutions remain well-behaved in the asymptotic region.


\section{The numerical solution}\label{SecV}

To conclude and achieve our objective, we will examine the numerical solutions of the BPS equations \eqref{Eq21} (or, equivalently, Eq. \eqref{Eq22}, since the topological sectors of $f$ and $g$ coincide). For the numerical analysis, we impose the topological boundary conditions specified in Eqs. \eqref{Eq13}, together with a numerical interpolation approach \footnote{The numerical interpolation approach consists of constructing an approximate representation of the radial profiles of the field variables $f(r)$ and $a(r)$ consistent with the regular asymptotic behavior \eqref{Eq13}. This procedure enables the determination of stable field profiles while simultaneously preserving regularity at the origin and the asymptotic decay required by the model’s topological structure. One can find more details on numerical interpolation in Refs. \cite{Burden,Butcher,Atkinson}.}. In the numerical procedure, a sufficiently large finite radial range, $r \in [0,100]$, is considered so that the profiles have already reached their vacuum values asymptotically at the upper limit. Since the point $r = 0$ is singular in polar coordinates, the integration is initiated in a small neighborhood of the origin, i.e., $r = \varepsilon \ll 1$, using the regular asymptotic expansions $a(r) \simeq n \mp r^2 + \cdots$ and $f(r) \simeq f_0\, r^{|n|} + \cdots$. This procedure ensures simultaneous compatibility with regularity at the core and with the solution's topological nature. The system is then integrated numerically, with the parameter $f_0$ adjusted so that the asymptotic condition $f(r) \to \pi$ is satisfied.

\begin{figure}[t]
    \centering
    \subfigure[Solution for the field variable $f(r)$.]{\includegraphics[width=7.5cm,height=6.5cm]{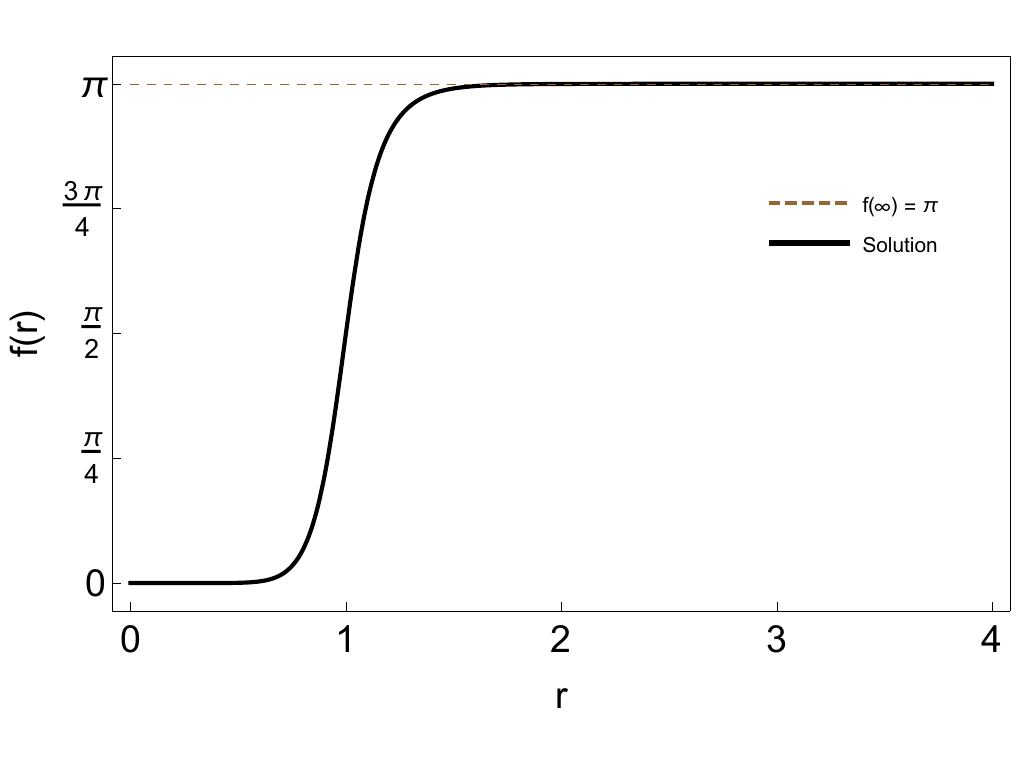}}\hfill
    \subfigure[Solution for the field variable $a(r)$.]{\includegraphics[width=7.5cm,height=6.5cm]{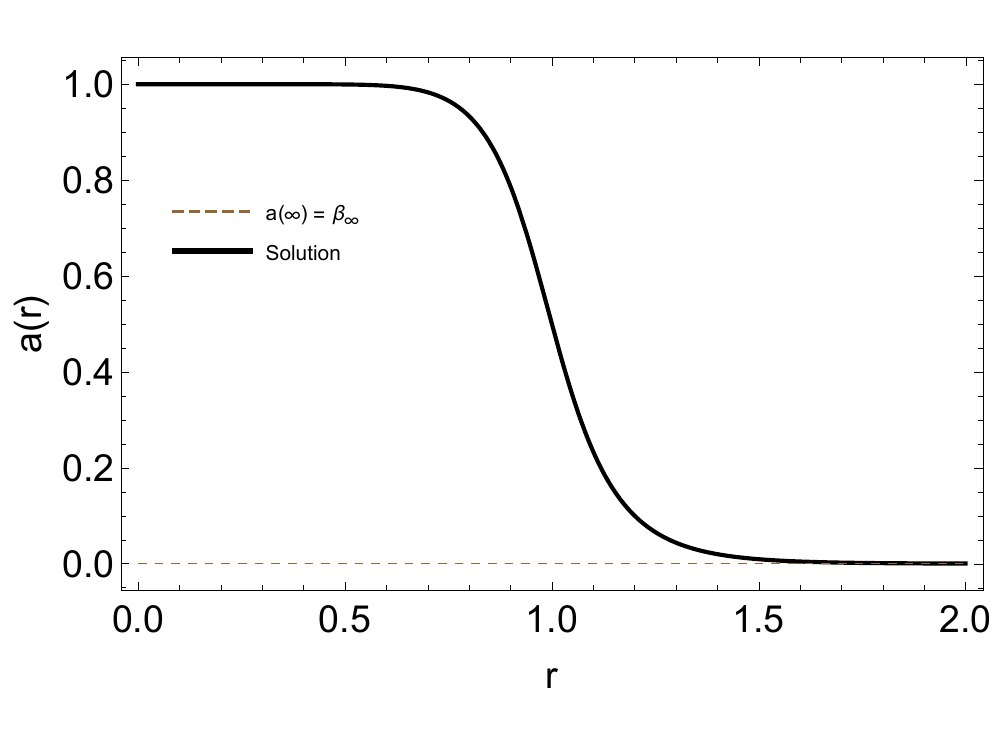}}
    \caption{Field profiles vs. radial coordinate. These solutions agree with the boundary conditions \eqref{Eq13}. This configurations have a BPS energy $\mathrm{E}_{\mathrm{BPS}}=\mp4\pi$, and an associated magnetic flux, viz., $\varphi_{\mathrm{flux}}=2\pi$. The solutions presented correspond to $n=e=1$ and $\beta_\infty=0$.}
    \label{Fig1}
\end{figure}

Implementing the aforementioned numerical procedure allows one to determine the profiles of the field variables $f(r)$, $g(r)$, and $a(r)$ \footnote{It is worth noting that $f\equiv g$.}. One exposes the resulting numerical solutions of the field variables in Figs. \ref{Fig1}(a) and \ref{Fig1}(b).

Examining the numerical results displayed in Figs. \ref{Fig1}(a) and \ref{Fig1}(b), one notes that the field variable $f(r)$ increases monotonically from the origin and approaches the asymptotic value $f(\infty) = \pi$, thereby describing a topological field configuration that interpolates between distinct points of the target manifold. This behavior confirms the existence of a regular solution near the vortex core, while the topological structure remains spatially localized. Conversely, the gauge field variable $a(r)$ starts from$ a(0)=n=1$ and decreases smoothly toward $a(\infty) = 0$. That result is consistent with the asymptotic analysis of the BPS sector, which ensures that the gauge field properly compensates the angular dependence of the sigma sector at spatial infinity.

\begin{figure}[!ht]
    \centering
    \subfigure[Magnetic field profile vs. $r$.]{\includegraphics[width=7.5cm,height=6cm]{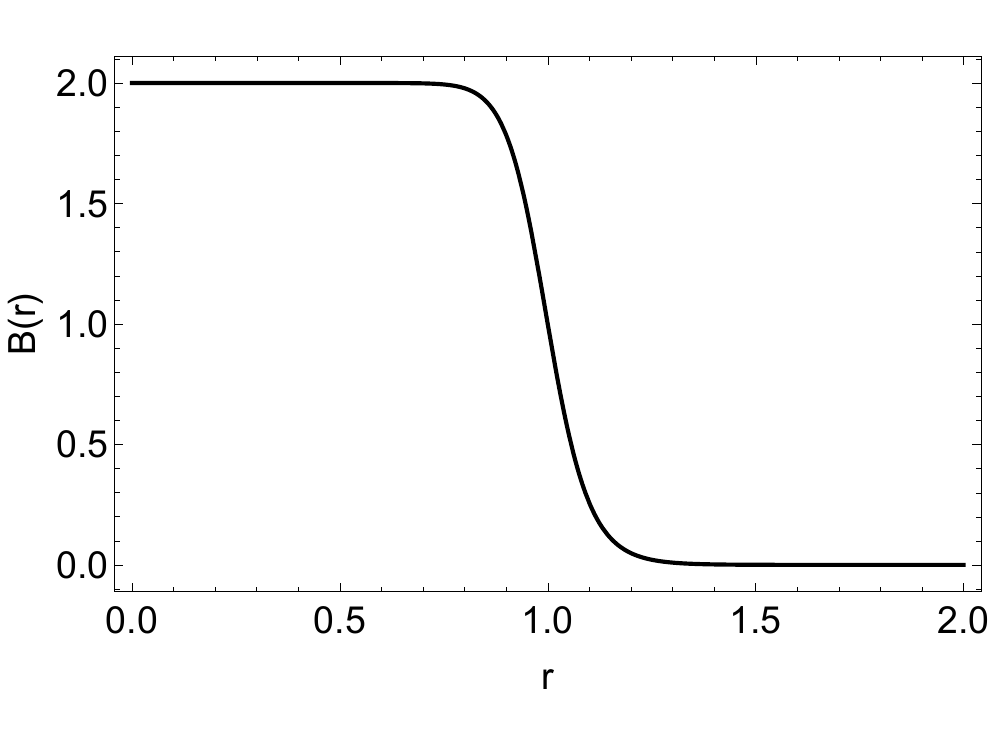}}\hfill
    \subfigure[Planar solution of the magnetic field.]{\includegraphics[width=7.5cm,height=6.5cm]{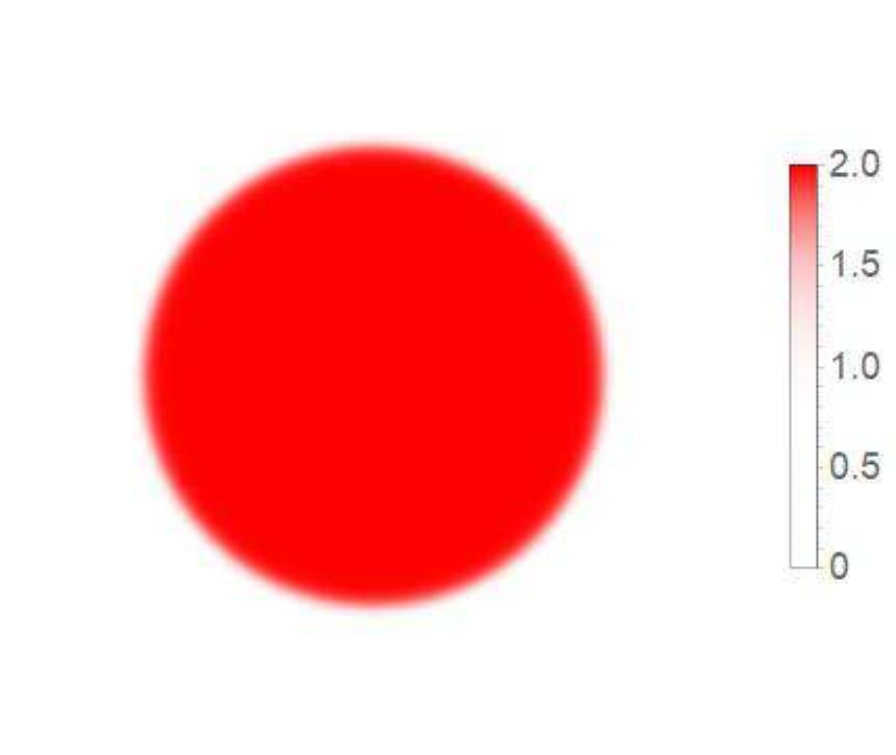}}
    \caption{Magnetic field generated by the vortex with flux $\varphi_{\mathrm{flux}}=2\pi$, for $n = e = 1$ and $\beta_\infty = 0$.}
    \label{Fig2}
\end{figure}

Once the numerical solutions for the field variables $f(r)$ and $a(r)$ have been obtained [Figs. \ref{Fig1}(a) and \ref{Fig1}(b)], we use the definition of the magnetic field $B(r) = -a'(r)/r$ and numerically examine the magnetic field profile of the topological vortex. By this means, the numerical profile of the magnetic field is obtained, as displayed in Figs. \ref{Fig2}(a) and \ref{Fig2}(b).

Inspection of the numerical results displayed in Figs. \ref{Fig2}(a) and \ref{Fig2}(b) show that the magnetic field is concentrated near the vortex core, becoming rapidly negligible outside a finite region. This behavior confirms the confinement of the magnetic flux. Furthermore, this profile ensures the existence of a vortex solution in the double-sigma model minimally coupled to the Maxwell field.

\begin{figure}[!ht]
    \centering
    \subfigure[BPS energy density vs. $r$.]{\includegraphics[width=7.5cm,height=6cm]{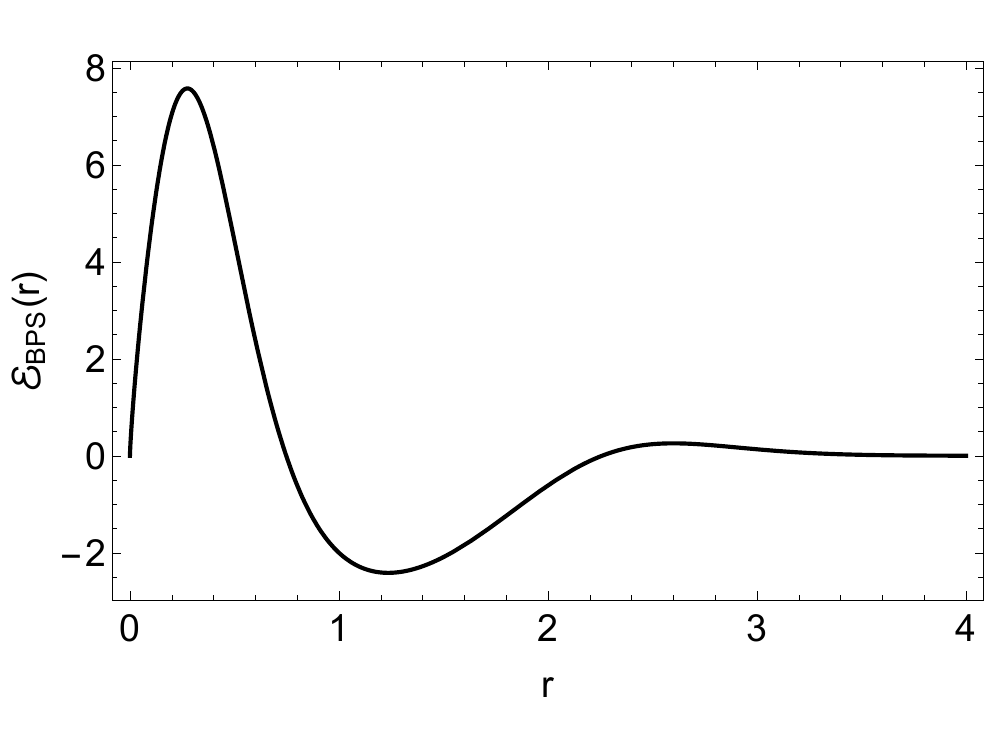}}\hfill
    \subfigure[Planar solution of the BPS energy density.]{\includegraphics[width=7cm,height=7cm]{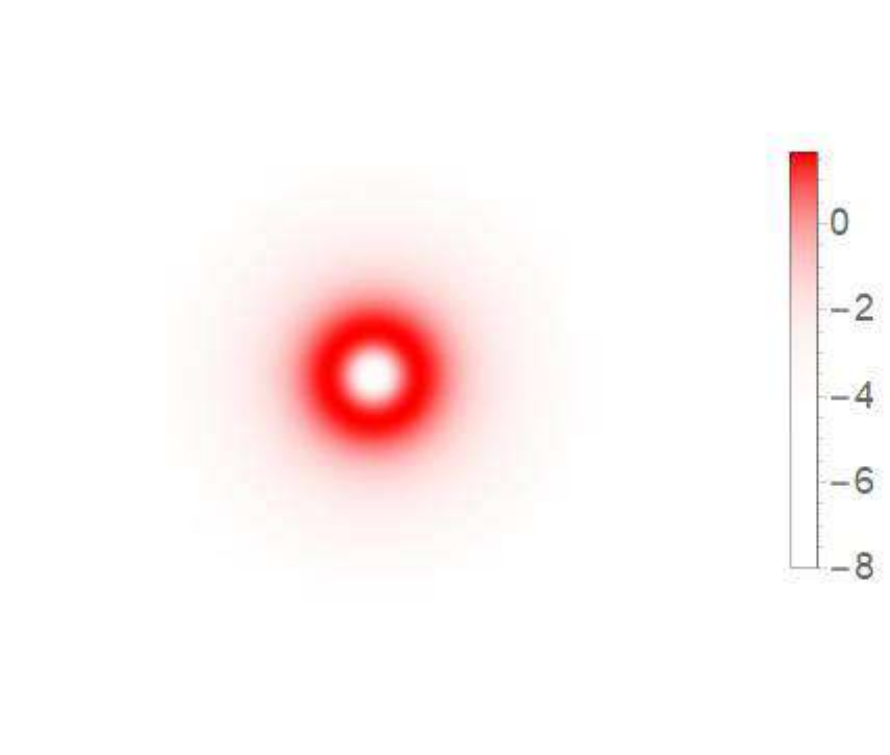}}
    \caption{BPS energy density concerning the magnetic vortex solution with total BPS energy $\mathrm{E}_{\mathrm{BPS}}=\pm2\pi$ and magnetic flux $\varphi_{\mathrm{flux}}=2\pi$, for $n=e=1$ and $\beta_\infty=0$.}
    \label{Fig3}
\end{figure}

Finally, one notes that the BPS energy density of the magnetic vortex is given in Eq. \eqref{Eq20}, or explicitly, $\mathcal{E}_{\mathrm{BPS}} = a'(2\cos f - 1) - 2a f'\sin f$. Thus, by employing the numerical solutions for the field variables $f(r)$ and $a(r)$ [see Figs. \ref{Fig1}(a) and \ref{Fig1}(b)] together with the definition of the BPS energy density, one obtains numerically the radial profile of the vortex BPS energy density in terms of r. We present the resulting profile in Figs. \ref{Fig3}(a) and \ref{Fig3}(b).

An analysis of the numerical results for the BPS energy density $\mathcal{E}_{\mathrm{BPS}}(r)$ reveals a ring-shaped, regular profile that is strongly localized around the vortex core and decays to zero as $r\to\infty$. This behavior ensures that the total energy remains finite and confirms that the numerical solution indeed saturates the BPS bound. In other words, the constructed configurations belong to the self-dual sector of the theory and describe the minimum-energy states compatible with the topological constraints imposed by the boundary conditions. From a physical viewpoint, the numerical results demonstrate that the model supports regular self-dual vortices with finite energy and quantized flux. Moreover, the coincidence between the profiles of the two sigma fields, $f(r)=g(r)$, indicates that both internal sectors act cooperatively in the BPS regime, effectively forming a single topological vortex structure. Meanwhile, the gauge field plays a crucial role in canceling the angular variation of the scalar fields, thereby stabilizing the solution and confining the magnetic flux near the vortex core.


\section{Summary and Conclusion}\label{SecVI}

In this work, we investigated the double O(3)-sigma model minimally coupled to a single Maxwell field in a three-dimensional spacetime. Within this framework, we examined the existence of BPS topological vortex solutions. The central motivation was to understand how two independent internal sectors, described by O(3)-sigma fields, interact through a single Abelian gauge field and how this interaction affects the formation, stability, and physical properties of the topological structures.

Our analysis showed that the presence of the coupling function $\mathcal{F}(\Theta)$ introduces, in principle, a nontrivial modulation of the dynamics of the $\Phi$-sector and the electromagnetic source itself. However, upon investigating the conditions for the existence of the BPS property, we found that self-duality necessarily requires the choice $\mathcal{F} = 1$ and $f(r) = g(r)$. Naturally, the first condition reduces the model to a sector in which both sigma fields contribute symmetrically to the effective vortex dynamics. Meanwhile, the condition $f(r) = g(r)$ shows that the two sigma sectors, although originally independent, become indistinguishable in the self-dual regime and effectively combine into a single topological structure.

The asymptotic analysis confirmed both the local and global consistency of the solutions. Near the origin, the sigma field vanishes with a power-law behavior governed by the winding number. In the asymptotic region, the linearization of the BPS equations showed that regularity of the solution requires $\beta_\infty = 0$, ensuring that the gauge field variable $a(r)$ decays at spatial infinity, which yields a quantized magnetic flux, i.e., $\varphi_\mathrm{flux}=2\pi n$. This result is particularly significant, as it demonstrates that the internal consistency of the self-dual regime dynamically selects the physically admissible behavior at spatial infinity.

Generally speaking, we concluded that the results obtained support the emergence of regular BPS vortex structures with finite energy and quantized magnetic flux. Moreover, within the BPS regime, the two internal sectors act cooperatively, producing a single effective topological configuration stabilized by the gauge field.


\section*{Acknowledgment}

F. C. E. Lima would like to express their sincere gratitude to the Conselho Nacional de Desenvolvimento Científico e Tecnológico (CNPq) and Fundação de Amparo \`{a} Pesquisa do Estado de S\~{a}o Paulo (FAPESP) for their valuable support. F. C. E. Lima is supported, respectively, for grants No. 171048/2023-7 (CNPq) and 2025/05176-7 (FAPESP). F. M. Belchior is suported for grants No. 151845/2025-5 (PD/CNPq).


\section*{Data availability}

No data was used for the research described in this article.

\bibliographystyle{apsrev4-2}
\bibliography{refs}


\section*{Appendix A - Consistency of the BPS approach}

Let us now inspect the consistency of the BPS equations \eqref{Eq21}. To this purpose, we demonstrate that the self-dual equations \eqref{Eq13} are equivalent to the equations of motion [\eqref{Eq7}–\eqref{Eq9}], taking into account the constraints of the $O(3)$-sigma model in both topological sectors, viz., $\Phi\cdot\Phi=1$ and $\Theta\cdot\Theta=1$.

\begin{center}
    \textbf{1. Reduction to the self-dual sector}
\end{center}

In the BPS sector, the completion of squares requires $\mathcal{F} = 1$, which implies $\mathcal{F}_\Theta = \partial_\mu \mathcal{F} = 0$. Accordingly, the equations of motion given in Eqs. [\eqref{Eq7}–\eqref{Eq8}] boils down to
\begin{align}
    D_\mu D^\mu\Phi+(D_\mu\Phi\cdot D^\mu\Phi)\Phi+V_\Phi-(\Phi\cdot V_\Phi)\Phi=0,
\label{Eq28}
\end{align}
and
\begin{align}
    D_\mu D^\mu\Theta+(D_\mu\Theta\cdot D^\mu\Theta)\Theta+V_\Theta-(\Theta\cdot V_\Theta)\Theta=0.
\label{Eq29}
\end{align}
Furthermore, taking into account that the dual potential, viz.,
\begin{align}
    V=\frac{1}{2}(\cos f+\cos g-1)^2,
\label{Eq30}
\end{align}
together with the ansätze introduced in Eqs. \eqref{Eq10} and \eqref{Eq11}, one finds that the equations of motion [\eqref{Eq7}–\eqref{Eq9}] reduce to a system defined purely in terms of radial variables, described by the profiles $f(r)$, $g(r)$, and $a(r)$. Accordingly, by writing the Euler–Lagrange equations of the theory, one obtains that
\begin{align}
    &f''+\frac{1}{r}f'-\frac{a^2}{r^2}\sin f\cos f+(\cos f+\cos g-1)\sin f=0,
\label{Eq31}
\end{align}
\begin{align}
    &g''+\frac{1}{r}g'-\frac{a^2}{r^2}\sin g\cos g+(\cos f+\cos g-1)\sin g=0.
\label{Eq32}
\end{align}

Meanwhile, the Maxwell equation [Eq. \eqref{Eq9}] takes the form
\begin{align}
        a''-\frac{1}{r}a'-a\left(\sin^2 f+\sin^2 g\right)=0.
    \label{Eq33}
\end{align}
Therefore, the second-order radial system in the self-dual sector boils down to the three equations \eqref{Eq31}, \eqref{Eq32}, and \eqref{Eq33}.

\begin{center}
    \textbf{2. Consistency of the self-dual equation for the sigma fields}
\end{center}

Taking into account the BPS equations \eqref{Eq21}, namely,
\begin{align}
    f'=\mp \frac{a}{r}\sin f, \qquad g'=\mp \frac{a}{r}\sin g, \qquad \mathrm{and} \qquad a'=\pm r(\cos f+\cos g-1).
\label{Eq34}
\end{align}
For simplicity, we restrict the demonstration to adopting the upper sign of the expressions \footnote{It is worth emphasizing that choosing the opposite sign leads to the same results.}
\begin{align}
    f'=- \frac{a}{r}\sin f,\qquad g'=- \frac{a}{r}\sin g, \qquad \mathrm{and} \qquad a'=r(\cos f+\cos g-1).
\label{Eq35}
\end{align}

Let us derive the expression $f'=\mp\frac{a}{r}\sin f$ with respect to the radial variable. That calculation leads to 
\begin{align}
    f''=-\left(\frac{a'}{r}-\frac{a}{r^2}\right)\sin f-\frac{a}{r}\cos f\,f'.
\label{Eq36}
\end{align}

By performing algebraic manipulations of Eq. \eqref{Eq36}, taking into account the self-dual equations \eqref{Eq34}, one finds
\begin{align}
    f''+\frac{1}{r}f'-\frac{a^2}{r^2}\sin f\cos f+(\cos f+\cos g-1)\sin f=0,
    \label{Eq39}
\end{align}
which coincides with the equation of motion announced in Eq. \eqref{Eq31}. Similarly, one finds that the equation $g'=\mp\frac{a}{r}\sin f$ satisfies the equation of motion \eqref{Eq34}.

\begin{center}
    \textbf{3. Consistency of the gauge field equation}
\end{center}

To conclude the discussion on the consistency of the self-dual equations \eqref{Eq21}, let us derive the gauge field equation from \eqref{Eq21}, viz., 
\begin{align}
    a''=(\cos f+\cos g-1)+r\frac{d}{dr}(\cos f+\cos g-1).
\label{Eq40}
\end{align}
i.e., 
\begin{align}
    a''=(\cos f+\cos g-1)-r\sin f\,f'-r\sin g\,g'.
    \label{Eq41}
\end{align}

Now, let us adopt the BPS equations for the sigma sectors \eqref{Eq34}, i.e., 
\begin{align}
    f'=-\frac{a}{r}\sin f \qquad \mathrm{and} \qquad g'=-\frac{a}{r}\sin g,
    \label{Eq42}
\end{align}
which lead us to 
\begin{align}
    -r\sin f\,f'=-r\sin f\left(-\frac{a}{r}\sin f\right)=a\sin^2 f,
    \label{Eq43}
\end{align}
and
\begin{align}
    -r\sin g\,g'=-r\sin g\left(-\frac{a}{r}\sin g\right)=a\sin^2 g.
    \label{Eq44}
\end{align}
Therefore, one can formulate Eq. \eqref{Eq41} as
\begin{align}
    a''=(\cos f+\cos g-1)+a(\sin^2 f+\sin^2 g).
    \label{Eq45}
\end{align}

By substituting Eq. \eqref{Eq42} into Eq. \eqref{Eq45}, one finds \begin{align}
    a''=\frac{a'}{r}+a(\sin^2 f+\sin^2 g),
    \label{Eq46}
\end{align}
i.e.,
\begin{align}
    a''-\frac{1}{r}a'-a(\sin^2 f+\sin^2 g)=0,
    \label{Eq47}
\end{align}
which coincides with the equation of motion exposed in Eq. \eqref{Eq35}. Consequently, the consistency of the three BPS equations of the theory presented in Eq. \eqref{Eq21} is verified.

\end{document}